\documentclass[a4paper]{article}

\usepackage{vmargin,amssymb,amsmath,epsfig,latexsym,booktabs,braket,
gensymb,amsbsy,stackrel,slashed}

\newcommand{\beq}{\begin{equation}}
\newcommand{\eeq}{\end{equation}}
\newcommand{\rar}{\, \rightarrow \,}
\newcommand{\cX}{{\cal X}}
\newcommand{\cY}{{\cal Y}}
\newcommand{\cZ}{{\cal Z}}

\newcommand{\vecy}[3]{ \left[ \!\!\! \begin{array}{c} #1 \\ #2 \\ #3 \end{array} \!\!\! \right]}

\newcommand{\sfrac}[2]{{\textstyle\frac{#1}{#2}}}
\newcommand{\half}{\sfrac{1}{2}}
\newcommand{\ihalf}{\sfrac{i}{2}}

\newcommand{\nln}{\nonumber\\}

\newcommand{\dK}{\delta K}
\newcommand{\sK}{\Sigma K}

\newcommand{\hopf}[1]{\mathrm{#1}}
\newcommand{\yang}{\hopf{Y}}

\newcommand{\alg}[1]{\mathfrak{#1}}

\newcommand{\gen}[1]{\mathbb{#1}{}}
\newcommand{\genyang}[1]{\widehat{\gen{#1}}{}}

\newcommand{\copro}{\mathrm{\Delta}}

\newcommand{\comm}[2]{[#1,#2]}
\newcommand{\acomm}[2]{\{#1,#2\}}
\newcommand{\gcomm}[2]{[#1,#2\}}

\begin{document}

\thispagestyle{empty}
\begin{flushright} 
TCDMATH--20--08
\end{flushright}

\vskip 5 cm

\begin{center}
{\huge \textbf{ Bound State Scattering Simplified}}

\vskip 2 cm

\textbf{M.~de Leeuw$^a$, B.~Eden$^b$, A.~Sfondrini$^{b,c}$}
\vskip 2 cm

\begin{minipage}{13.1 cm}
\noindent $^a$School of Mathematics \& Hamilton Mathematics Institute, Trinity College Dublin, \\ \phantom{$^a$}20 Westland Row, Dublin 2, Ireland, {\it E-mail:} mdeleeuw@maths.tcd.ie \\
$^b$Institut f\"ur theoretische Physik, ETH Z\"urich,  Wolfgang-Pauli-Strasse 27, 8093 Z\"urich,\\ \phantom{$^b$}Switzerland, {\it E-mail:} bueden@phys.ethz.ch\\
$^c$Dipartimento di Fisica e Astronomia, Universit\`a degli Studi di Padova,\\
\phantom{$^c$}\& Istituto Nazionale di Fisica Nucleare, Sezione di Padova,\\
\phantom{$^c$}via Marzolo 8, 35131 Padova, Italy. \textit{E-mail:} alessandro.sfondrini@unipd.it
\end{minipage} 
 
\end{center}

\vskip 3 cm

In the description of the AdS$_5$/CFT$_4$ duality by an integrable system the scattering matrix for bound states plays a crucial r\^ole: it was initially constructed for the evaluation of finite size corrections to the planar spectrum of energy levels/anomalous dimensions by the thermodynamic Bethe ansatz, and more recently it re-appeared in the context of the glueing prescription of the hexagon approach to higher-point functions.
In this work we present a simplified form of this scattering matrix and we make its pole structure manifest.  We  find some new relations between its matrix elements and also present an explicit form for its inverse. We finally discuss some of its properties including crossing symmetry. 
%
Our results will hopefully  be useful for TBA applications, in simplifying the complicated sum-integrals arising from the glueing of hexagons as well as help towards understanding universal features of the AdS$_5$/CFT$_4$ scattering matrix. 

\newpage

\section{Introduction}

In the study of the AdS$_5$/CFT$_4$ correspondence \cite{123} the problem of computing string energy levels or, in the dual ${\cal N} \, = \, 4$ super Yang-Mills theory (SYM), the planar anomalous dimensions of gauge-invariant composite operators has been related to an integrable system, namely an extended and deformed version of the Heisenberg spin chain \cite{beiStau}. The form of the $S$-matrix governing the scattering of the excitations on this chain is constrained by symmetry \cite{beisertPSU22} up to one overall phase \cite{BES}. 

This integrable model is able to provide all orders in the 't Hooft coupling $\lambda$ in the asymptotic regime of infinite spin chain length. Finite size corrections have been addressed by L\"uscher corrections~\cite{luscher} first and then, systematically, by the thermodynamic Bethe ansatz (TBA) \cite{TBA} which requires taking into account the bound states of the theory. An $S$-matrix for such bound states generalising \cite{beisertPSU22} was first derived in \cite{Arutyunov:2008zt} for bound states up to length two and then extended to arbitrary bound states in \cite{glebBound} on grounds of Lie algebra and Yangian symmetry \cite{niklasYang}. It has a block diagonal structure with two equal $1 \times 1$ blocks called $\cX$, two equal $4 \times 4$ blocks $\cY$ and finally a $6 \times 6$ block named $\cZ$. In the original work \cite{glebBound}, $\cX$ is given explicitly --- it is essentially a generalised hypergeometric $_4F_3$ function --- but the other blocks were only implicitly defined involving matrix inverses that seemed hard to simplify. This in principle poses an obstacle to L\"uscher-type computations which rely on the explicit form of the $S$-matrix.

Recently, the computation of three-point functions in ${\cal N} \, = \, 4$ SYM became accessible to ``integrability'' methods by the invention of the hexagon approach \cite{BKV}. Here one cuts the closed string world sheet into two hexagonal patches; the gauge theory equivalent is cutting up Feynman diagrams on the sphere into two halves. To obtain the full quantum result these patches have to be glued together again \cite{BKV} by inserting complete sets of bound states on the edges. Hence also in this context the scattering of bound states is of prime importance.

Finally, higher point functions can apparently be computed by hexagon tessellations, using the hexagon operator of the three-point problem as an elementary patch and glueing appropriately \cite{cushions,shotaThiago1,positivity}. At weak coupling, the procedure is technically involved already at one loop not at last because of the complexity of the bound state $S$-matrix needed in the glueing. Yet, in a recent attempt \cite{fivePt} on verifying and extending existing work at five points \cite{shotaThiago2} we noticed that the bound state $S$-matrix had to be a much simpler object than the original work \cite{glebBound} suggested. In this work we tackle the programme of simplifying the matrix. We are able to provide a completely explicit writing in terms of relatively concise objects. Moreover, we uncover some new structure between the elements of the bound state $S$-matrix.

The note has the following structure: first, we recall the basic construction and results of \cite{glebBound}. After this we discuss our approach to simplifying the bound state $S$-matrix and give compact expression for its components. Their pole structure is clear from our new expressions. Finally we discuss some discrete symmetries of the $S$-matrix and crossing symmetry.

\section{Review of bound state scattering}

Let us briefly review the construction of the bound-state $S$-matrix as presented in \cite{glebBound}. The two-particle $S$-matrix $S_{12}$ has to commute with the symmetry of the problem:
\begin{align}
\gen{J}_{21}  S_{12} = S_{12}   \gen{J}_{12},
\end{align}
Here $\gen J$, the manifest symmetry of the S-matrix, spans a subalgebra of the superconformal algebra~$\mathfrak{psu}(2,2|4)$ given by two-copies of $\mathfrak{su}(2|2)$; moreover, and crucially, this algebra is centrally extended as discovered in \cite{beisertPSU22} (see also~\cite{Arutyunov:2009ga} from a derivation of the central extension from the string worldsheet). Hence the algebra of interest will be the centrally extended $\mathfrak{su}(2|2)$ of~\cite{beisertPSU22}.

\paragraph{Lie Superalgebra.}


There are two $\alg{su}(2)$'s,
spanned by the generators $\gen{L}^a{}_b,\gen{\tilde L}^\alpha{}_\beta$ with 
$\gen{L}^a{}_a=\gen{\tilde L}^\alpha{}_\alpha=0$,
two sets of supercharges $\gen{Q}^\alpha{}_b,\gen{\bar Q}^a{}_\beta$
and three central elements $\gen{H},\gen{C},\gen{\bar C}$.
Latin letters $a,b,\ldots=1,2$ run over the Grassmann even indices and
Greek letters $\alpha,\beta,\ldots=1,2$ run over the odd indices.
The non-trivial commutation relations are given by
\begin{align}\label{eq:DefExtsu22}
\comm{\gen{L}^a{}_b}{\gen{L}^c{}_d} &=
\delta^c_b \gen{L}^a{}_d - \delta^a_d \gen{L}^c{}_b ,
&
\comm{\gen{\tilde L}^\alpha{}_\beta}{\gen{\tilde L}^\gamma{}_\delta} &=
 \delta^\gamma_\beta\gen{\tilde L}^\alpha{}_\delta
- \delta^\alpha_\delta \gen{\tilde L}^\gamma{}_\beta,
\nln
\comm{\gen{L}^a{}_b}{\gen{Q}^\alpha{}_c} &=
-\delta^a_c \gen{Q}^\alpha{}_b + \half\delta^a_b \gen{Q}^\alpha{}_c,
&
\comm{\gen{\tilde L}^\alpha{}_\beta}{\gen{Q}^\gamma{}_a} &=
\delta^\gamma_\beta \gen{Q}^\alpha{}_a
- \half\delta^\alpha_\beta\gen{Q}^\gamma{}_a ,
\nln
\comm{\gen{L}^a{}_b}{\gen{\bar Q}^c{}_\alpha} &=
 \delta^c_b \gen{\bar Q}^a{}_\alpha - \half\delta^a_b \gen{\bar Q}^c{}_\alpha,
&
\comm{\gen{\tilde L}^\alpha{}_\beta}{\gen{\bar Q}^a{}_\gamma} &=
-\delta^\alpha_\gamma \gen{\bar Q}^a{}_\beta + \half\delta^\alpha_\beta\gen{\bar Q}^a{}_\gamma ,
\nln
\acomm{\gen{Q}^\alpha{}_a}{\gen{Q}^\beta{}_b} &=
\varepsilon_{ab}\varepsilon^{\alpha\beta} \gen{C},
&
\acomm{\gen{\bar Q}^a{}_\alpha}{\gen{\bar Q}^b{}_\beta} &=
 \varepsilon^{ab}\varepsilon_{\alpha\beta} \gen{\bar C},
\nln
\acomm{\gen{Q}^\alpha{}_a}{\gen{\bar Q}^b{}_\beta} &=
\delta_a^b\gen{\tilde L}^\alpha{}_\beta + \delta_\beta^\alpha \gen{L}^b{}_a
+ \half\delta_\beta^\alpha\delta_a^b\gen{H}.
\end{align}
By setting $\gen{C},\gen{\bar C}=0$ the algebra reduces to $\alg{sl}(2|2)$. \\


Yet, the key ingredient of the construction in \cite{glebBound} is the Yangian of the centrally extended $\mathfrak{su}(2|2)$ \cite{niklasYang}:

\paragraph{Hopf Algebra.}

The Hopf algebra structure depends on a central element $\gen{U}$ which is called braiding element. It is used to deform the coproduct of the Lie generators $\gen{J}$ in the following way
\begin{align}
\label{eq:coalgbraid}
\copro(\gen{J}) &= \gen{J} \otimes 1 + \gen{U}^{[\gen{J}]}\otimes \gen{J},
\end{align}
where the weight $[\gen{J}]$ is defined by 
$[\gen{L}]=[\gen{\tilde L}]=[\gen{H}]=0$, $[\gen{Q}]=-[\gen{\bar Q}]=1$
and $[\gen{C}]=-[\gen{\bar C}]=2$.

By requiring that the coproduct of the central elements
is cocommutative, one can derive a relation between the
braiding element and the central elements.
\begin{equation}\label{eq:CviaU}
\gen{C} = \ihalf g(\gen{U}^2-1), \qquad \gen{\bar C}
 = \ihalf g(1-\gen{U}^{-2}).
\end{equation}
%

\paragraph{Extended Yangian}\label{sec:YangianSL22}

In addition to the above elements $\gen{J}^I,\gen{U}\in\hopf{A}$, 
the Yangian algebra $\yang$ is generated by level-one elements $\genyang{J}^I$. 
They obey the conventional Yangian relations
\[
\gcomm{\gen{J}^I}{\genyang{J}^J} = f^{IJ}{}_K \genyang{J}^K,
\]
with the structure constants $f^{IJ}{}_K$.
The only non-trivial part of the Hopf algebra is the coproduct,
since the remaining Hopf algebra structures are readily derived from it.
\[
\copro(\genyang{J}^I) = \genyang{J}^I \otimes 1 + \gen{U}^{[I]}\otimes \genyang{J}
+
 (-1)^{|J||K|}\half\hbar f^I{}_{JK}
\gen{J}^J\gen{U}^{[K]}\otimes \gen{J}^K
,
\]
Let us spell out the coproduct of the supercharges $\genyang{Q}^\alpha{}_a$,
since the rest follows by using the commutation relations
\begin{align}
\copro(\genyang{Q}^\alpha{}_a) =\mathord{}&
  \genyang{Q}^\alpha{}_a\otimes1
+ \gen{U}\otimes \genyang{Q}^\alpha{}_a
+  \frac{\hbar}{2}
 [
\gen{Q}^\alpha{}_c\otimes\gen{L}^c{}_a - \gen{L}^c{}_a\gen{U}\otimes\gen{Q}^\alpha{}_c +
\gen{Q}^\gamma_a\otimes\gen{\tilde L}^\alpha_\gamma
- \gen{\tilde L}^\alpha_\gamma\gen{U}\otimes\gen{Q}^\gamma_a   
\nonumber\\
&\, 
-\varepsilon^{\alpha\beta}\varepsilon_{ab}\gen{\bar Q}^b{}_\beta\otimes\gen{C}
+ \varepsilon^{\alpha\beta}\varepsilon_{ab}\gen{C}\gen{U}^{-1}\otimes\gen{\bar Q}^b{}_\beta
+\half\gen{Q}^\alpha{}_a\otimes\gen{H}
- \half\gen{H}\gen{U}\otimes\gen{Q}^\alpha{}_a
  ].
\end{align}
The coupling constant $g$  also takes the role of the deformation parameter in the definition of the Yangian. \\

The bound state $S$-matrix is then by definition the invertible operator that intertwines the usual and opposite coproduct
\begin{align}
\Delta^{op} (\gen{J}) \ S_{12} = S_{12} \ \Delta (\gen{J}),\qquad
\Delta^{op}=\Delta\circ\Pi^g,
\end{align}
for any generator $\gen{J}$ of the Yangian of centrally extended $\mathfrak{su}(2|2)$ in the corresponding representation. Here the opposite co-product is defined by means of the graded permutation operator~$\Pi^g$.

\subsection{Bound state representation}

The $K$-particle bound state representation is spanned by vectors that can be identified with monomials in variables $\theta_{1,2}$ and $w_{1,2}$
\begin{align}
|n_1,n_2,n_3,n_4\rangle = \theta_1^{n_1}\theta_2^{n_2}w_1^{n_3}w_2^{n_4},
\end{align}
such that $K = n_1+n_2+n_3+n_4$. The variables $\theta_{3,4}$ are odd while the variables $w_{1,2}$ are even. Consequently we have $0\leq n_{1,2}\leq 1$. The dual basis is given by 
\begin{align}
\langle n_1,n_2,n_3,n_4| = \partial_{w_2}^{n_4}\partial_{w_1}^{n_3}\partial_{\theta_2}^{n_2}\partial_{\theta_1}^{n_1},
\end{align}
such that the inner product is
\begin{align}
\langle m_1,m_2,m_3,m_4|n_1,n_2,n_3,n_4\rangle = \delta_{m_1,n_1}\delta_{m_2,n_2}\delta_{m_3,n_3}\delta_{m_4,n_4} n_1!n_2!n_3!n_4!.
\end{align}
The algebra generators of centrally extended $\alg{su}(2|2)$ are then represented by differential operators of the form
\begin{align}
&L^a_b = w_b \partial_{w_a} -\half \delta^a_b w_c \partial_{w_c},
&&\tilde{L}^\alpha_\beta = \theta_\beta \partial_{\theta_\alpha} - \half \delta^\alpha_\beta \theta_\gamma \partial_{\theta_\gamma},\\
&Q^\alpha_b = a\, w_b \partial_{\theta_\alpha} + b\,\epsilon_{ba}\epsilon^{\alpha\beta} \theta_\alpha \partial_{w_b},
&&\bar{Q}^a_\beta = c\, \theta_\beta \partial_{w_a} + d\,  \epsilon^{ab}\epsilon_{\beta\alpha} w_b \partial_{\theta_\alpha}
\end{align}
The supersymmetry generators depend on four parameters $a,b,c,d$ that are parameterized as
\begin{align}
&a  = \sqrt{\frac{g}{2K}}\gamma && b= -\sqrt{\frac{g}{2K}} \frac{i}{\gamma}\Big(1-\frac{x^+}{x^-}\Big) \\
&c  = -\sqrt{\frac{g}{2K}}\frac{\gamma}{x^+} && d = \sqrt{\frac{g}{2K}}\frac{x^+-x^-}{i\gamma}.
\end{align}
The variables $x^\pm$ satisfy the following relations
\begin{align}
x^+ + \frac{1}{x^+} - x^- -\frac{1}{x^-} = \frac{2iK}{g}\,,\qquad
\frac{x^+}{x^-}=e^{ip}.
\end{align}
The representation parameter $\gamma$ arbitrary as it can be changed by rescaling single-particle states, and our results will hold for general $\gamma$. It is convenient to choose
\begin{align}
\label{eq:gamma}
\gamma = \sqrt{i(x^- - x^+)\,U}\,,
\end{align}
which makes the representation unitary and provides it with nice analytic properties~\cite{Arutyunov:2009ga}. (Multiplying $\gamma$ in \eqref{eq:gamma} by a function $e^{i\phi(p)}$ such that $\phi(p)$ is a real analytic function also yields a unitary representation.)
Let us also introduce the rapidity $u$ and the rescaled rapidity $v$
\begin{equation}
u = \frac{1}{2} \bigg ( x^+ + \frac{1}{x^+}  + x^- +\frac{1}{x^-} \bigg) \, , \qquad v = -\frac{ig}{2} u \, .
\end{equation}
The braiding factor is given by $\gen{U} = \sqrt{{x^+}/{x^-}}$.

\subsection{Two-particle basis}

The two-particle $S$-matrix scatters states of the form $|m_1,m_2,m_3,m_4\rangle\otimes |n_1,n_2,n_3,n_4\rangle $. We will use the convention that states from space one are labelled by integers $K_1,k,m$ and states from space two are labelled by $K_2,l,n$. Moreover, it is convenient to introduce 
\begin{align}\label{defs}
&\bar{k} = K_1 - k -1,
&&\bar{l} = K_2 - l -1,
&&\bar{m} = K_1 - m -1,
&&\bar{n} = K_2 - n -1,
&& \Sigma K = \frac{K_1+K_2}{2}.
\end{align}
We only need to restrict to the eigenspaces $V_{r,\ell}$ of $\Delta \gen{L}^1_1$ and $\Delta \tilde{\gen{L}}^1_1$. The eigenvalues of  $\Delta \tilde{\gen{L}}^1_1$ take values $r = \pm1,\pm\half,0$, while $\Delta\gen{ L}^1_1$ has eigenvalues $\ell = -\Sigma K,\ldots, \Sigma K$. Let us label the vectors that span these eigenstates by their eigenvalues under these operators. 

\paragraph{Case I} First, let us look at the vector space where $r=\pm 1$
\begin{align}
V_{1,\ell} 
&=\Big\{ |1,0,k,\bar{k}\rangle\otimes |1,0,l,\bar{l}\rangle \Big\} _{2\ell = k+l-\bar{k}-\bar{l}}=: 
\Big\{ |k,l\rangle^{(1)}   \Big\}  \\
V_{-1,\ell} 
&=\Big\{ |0,1,k,\bar{k}\rangle\otimes |0,1,l,\bar{l}\rangle \Big\} _{2\ell = k+l-\bar{k}-\bar{l}}=: 
\Big\{ |k,l\rangle^{(-1)}   \Big\} 
\end{align}
We only need to label the vectors $|k,l\rangle^{(\pm1)}$ by their $\Delta \tilde{\gen{L}}^1_1$ eigenvalue since the eigenvalue of $\Delta \gen{L}^1_1$ can be directly read off from the labels $k,l$ in the state. The labels $k,l$ take the values $k=0,\ldots K_1-1$ and  $l=0,\ldots K_2-1$.
In \cite{glebBound} these vectors were labeled by $\mathrm{IA},\mathrm{IB}$ respectively.

\paragraph{Case II} 

Second, we consider the subspaces with eigenvalues $r = \pm\half$. The basis vectors of $V_{r,\ell}$ are
\begin{align}
V_{\half,\ell} &= \Big\{
|1,0,k, \bar{k}\rangle\otimes |0,0,l, \bar{l}+1\rangle,
 |0,0,k, \bar{k}+1\rangle\otimes |1,0,l, \bar{l}\rangle,\nln
&\qquad|1,0,k, \bar{k}\rangle\otimes |1,1,l-1,\bar{l}\rangle,
|1,1,k-1,\bar{k}\rangle\otimes |1,0,l, \bar{l}\rangle 
\Big\} \\
&=: \Big\{ |k,l\rangle^{(\half)}_1 ,|k,l\rangle^{(\half)}_2 ,|k,l\rangle^{(\half)}_3 ,|k,l\rangle^{(\half)}_4 \Big\}
\end{align}
where $\ell  = k+l - \Sigma K -\half$. We define $V_{-\half,\ell}$ analogously. 

\paragraph{Case III} 

Finally, there is the case when $r=0$. Here, the basis vectors are
\begin{align}
V_{0,\ell}  &= \{
 |0,0,k,\bar{k}+1\rangle\otimes |0,0,l,\bar{l}+1\rangle,
|0,0,k,\bar{k}+1\rangle\otimes |1,1,l-1,\bar{l}\rangle,
 |1,1,\bar{k},k-1\rangle\otimes |0,0,\bar{l}+1,l\rangle,\nln
&\qquad  |1,1,\bar{k},k-1\rangle\otimes |1,1,\bar{l},l-1\rangle 
 |1,0,\bar{k},k\rangle\otimes |0,1,\bar{l}+1,l-1\rangle  |0,1,\bar{k}+1,k-1\rangle\otimes |1,0,\bar{l},l\rangle 
\Big\} \\
&=: \Big\{ |k,l\rangle^{(0)}_1 ,|k,l\rangle^{(0)}_2 ,|k,l\rangle^{(0)}_3 ,|k,l\rangle^{(0)}_4 ,|k,l\rangle^{(0)}_5 ,|k,l\rangle^{(0)}_6 \Big\}
\end{align}
Vectors from the different cases can be mapped to each other by using the supersymmetry generators. This can be exploited to compute the bound state $S$-matrix from its defining intertwining property.

\subsection{$S$-matrix}

The $S$-matrix is defined up to a normalisation factor. We choose our $S$-matrix to be normalized such that
\begin{align}
S \cdot |0,0\rangle^{(\pm1)} = |0,0\rangle^{(\pm1)}.
\end{align}
\textit{i.e.} the scattering of highest weight fermionic states has eigenvalue one.

\paragraph{Case I $S$-matrix}

The $S$-matrix restricted to the subspaces $V_{\pm1}$ takes the simple form
\begin{align}
S \cdot |k,l\rangle^{(\pm1)} = \sum_{n=0}^{k+l} \mathcal{X}^{kl}_n(v) |n,k+l-n\rangle^{(\pm1)},
\end{align}
where $v = v_1-v_2$ and 
\begin{align}\label{eq:SmatX}
 \mathcal{X}^{kl}_n =&
 \frac{\Gamma(K_2-l)}{\Gamma(K_1-n)}
 \frac{\Gamma(v + \dK +1)}{\Gamma(v - \dK+1-l+n)} 
 \frac{\Gamma(v+\sK-k-l)}{\Gamma(v+\sK)}\times
\nonumber\\ 
&\times\sum_{q=0}^{k} \binom{k}{k-q}\binom{l}{n-q} \frac{\Gamma(K_1-q)}{\Gamma(K_2-k-l+q)}\frac{\Gamma(v - \dK+1+q)}{\Gamma(v - \dK+1-q)}.
\end{align}
Notice that $\cX$ is purely of difference form and actually coincides with the $\mathfrak{su}(2)$ universal $R$-matrix evaluated in the symmetric representations \cite{Arutyunov:2009ce}.

\paragraph{Case II $S$-matrix}

The $S$-matrix that describes the scattering of the fermionic states from the subspaces $V_{\pm1/2}$ can be obtained by using the supersymmetry generators. By using Yangian symmetry it is possible to define four operators that relate the four basis vectors $|k,l\rangle^{(1/2)}_i$ to the vector $|k,l\rangle^{(1)}$. In this way, one can express the matrix $\cY$ in terms of $\cX$ as follows
\begin{align}\label{defY}
A \cY^{kl}_n = B \cX^{kl}_n +  B _+ \cX^{k+1,l-1}_n + B_- \cX^{k-1,l+1}_n,
\end{align}
where $A,A_\pm$ are $4\times4$ matrices with some rather involved components whose explicit form can be found in \cite{glebBound}.

\paragraph{Case III $S$-matrix}
By similar arguments, the $S$-matrix restricted to $V_{0}$  can be obtained from $\cY^{kl}_n$ with various shifted indices 
\begin{align}\label{defZ}
C \cZ^{kl}_n = D \cY^{kl}_n +  D_1 \cY^{k,l-1}_n+  D_2 \cY^{k-1,l}_{n}+  D_3 \cY^{k-1,l}_{n-1}+  D_4 \cY^{k,l-1}_{n-1}.
\end{align}
We again encounter complicated matrix inversion and multiplication.

\section{Simplifying $\cX$}

Formula \eqref{eq:SmatX} for the $\cX$-matrix is rather concise. However, this or any other writing conceals the pole structure of the object, which is essential knowledge for example for residue calculations arising from the glueing procedure of the hexagon approach \cite{shotaThiago2,fivePt}. Furthermore, $\cX$ is a hypergeometric function and hence it obeys a number of contiguity equations. 

 
\paragraph{Pole decomposition} From the explicit expression \eqref{eq:SmatX}, it is easy to see that the only poles are at
\begin{align}
&v =- \Sigma K - \alpha, && \alpha=0,\ldots, k+l,
\end{align}
which are all in the complex plane. These are simple poles, as we can make apparent in the following elegant decomposition
\begin{align}
\cX^{kl}_n (v) = \delta^k_n + \frac{k!l!}{m!n!} \sum_{\alpha=0}^{k+l} (-1)^{l-n-\alpha} \frac{k+l-\alpha}{v+\Sigma K - k-l+\alpha}\sum_{\beta = 0}^\alpha \binom{m}{\beta}
 \binom{n}{\alpha - \beta} \binom{\bar{k} + \beta}{l}\binom{\bar{l} +\alpha- \beta}{k}.
 \end{align}
The second sum can actually be performed and gives an expression in terms of ${}_4F_3$.

\paragraph{Recursion relations} From Yangian symmetry it can be shown that $\cX$ satisfies the recursion relations \cite{fivePt} 
\begin{align}
    \cX^{k+1,l}_n &=\frac{1}{\bar{k}} \bigg[\frac{(\bar{n}-k) \bar{m}}{v +\Sigma K-k-l-1} \cX^{k,l}_n + 
    \frac{(\bar{n}+1) (v+ \delta K +l-n+1)}{v + \Sigma K-k-l-1} \cX^{k,l}_{n-1}, \bigg] \, ,\label{eq:Xkplus}\\
    \cX^{k-1,l}_n &= \frac{1}{k} \bigg[\frac{(n-\bar{k}) m}{v - \Sigma K+k+l+1} \cX^{k,l}_n + \frac{(n+1) (v - \delta K -l+n+1)}{v - \Sigma K+k+l+1} \cX^{k,l}_{n+1}, \bigg] \, ,\label{eq:Xkmin}\\
    \cX^{k,l+1}_n &= \frac{1}{\bar{l}}\bigg[ \frac{\bar{m} (v -\delta K -l+n)}{v +\Sigma K-k-l-1} \cX^{k,l}_n + \frac{(\bar{n}+1) (\bar{m}-l-1)}{v +\delta K-k-l-1} \cX^{k,l}_{n-1} \bigg] \, ,\label{eq:Xlplus}\\
    \cX^{k,l-1}_n &= \frac{1}{l}\bigg[ \frac{m (v + \delta K + l-n)}{v -\Sigma K+k+l+1} \cX^{k,l}_n + \frac{(n+1) (m-\bar{l}-1)}{v -\delta K+k+l+1} \cX^{k,l}_{n+1} \bigg]\label{eq:Xlmin}.
\end{align}
Here $\bar k$ etc are defined in \eqref{defs}.
Notice that the cases $\cX^{k\pm1,l}_n$ and $\cX^{k,l\pm1}_n$ are each related by switching barred and unbarred indices. From these relations we see that for fixed $k,l$ all $\cX$-matrices with shifted indices can be brought into a standard form $\cX^{k,l}_n,\cX^{k,l}_{n\pm1}, \ldots$. 

Finally, by successively using \eqref{eq:Xkplus} and \eqref{eq:Xkmin} we obtain the following 
\begin{align}
    \cX^{k,l}_{n-1} =& \frac{(n+1) (\bar{m}+1) (v -\delta K-l+n+1)}{(\bar{n}+1) (m+1) (v +\delta K+l-n+1)}\cX^{k,l}_{n+1} + \nonumber\\
    &~\Bigg[\frac{(n+1) \bar{n} (v + \delta K +l-n)(v-\delta K-l+n+1)}{(\bar{n}+1) (m+1) (k-\bar{n}+1)(v +\delta K +l-n+1)} + \nonumber\\
    &\quad+\frac{\bar{m} (k-\bar{n})}{(\bar{n}+1) (v +\delta K +l-n+1)}\\
&   \quad-\frac{(k+1) \bar{k} (v -\Sigma K+k+l+2) (v +\Sigma K-k-l -1)}{(\bar{n}+1)
   (m+1) (k-\bar{n}+1) (v +\delta K +l-n+1)}\Bigg]\cX^{k,l}_n \nonumber
\end{align}
From this we can also remove any $\cX$-matrix whose $n$ index is shifted by a negative integer. This can now be used to compare different, possibly equivalent ways of writing the other entries of the bound state $S$-matrix.

Moreover, we can recursively construct the $\cX$-matrix starting from $\cX^{00}_0=1$ and raising the indices by making use of these relations. For instance, using \eqref{eq:Xkplus} we find
\begin{align}
 \cX^{10}_0 = 
 \frac{K_2-1}{v+\sK -1} \cX^{00}_0 + 
 \frac{K_1}{K_1-1}\frac{v+\dK+1}{v+\sK-1} \cX^{00}_{-1}.
\end{align}
Since $\cX^{00}_{-1}=0$, we find
\begin{align}
 \cX^{10}_0 = 
 \frac{K_2-1}{v+\sK -1} .
\end{align}
By repeating this argument we can derive any $\cX^{kl}_0$. We can then use, for example \eqref{eq:Xkmin}, to compute $\cX^{kl}_1$ in terms of $\cX^{kl}_0$ and $\cX^{k-1,l}_0$ and work from there to general $n$.

\paragraph{Useful identities}

It is clear that swapping barred and unbarred indices should leave $\cX$ invariant and indeed
\begin{align}
\cX^{kl}_n = \cX^{\bar{k}\bar{l}}_{\bar{n}}.
\end{align}
We also have the symmetry property
\begin{align}\label{eq:SymX}
\frac{\cX^{kl}_n}{{}^{(1)}\langle k,l| k,l\rangle^{(1)}}
=
\frac{\cX^{nm}_k}{{}^{(1)}\langle n,m| n,m\rangle^{(1)}}.
\end{align}
We note that the inverse of $\cX(v)$ is simply given by $\cX(-v)$, \textit{i.e.}
\begin{align}
\sum_{a=0}^{k+l} \cX^{kl}_a(v) \cX^{a,k+l-a}_n(-v) = \delta_{kn}.
\end{align}
In what follows we will use the identities
\begin{align}
 (k-n) ( v+\Sigma K-k-l) \cX^{kl}_n - k  (K_2-m) \cX^{k-1,l}_n + l (K_1-n)\cX^{k,l-1}_{n-1}  &= 0, \label{eq:Xiden1}\\
 (n-k) ( v-\Sigma K+k+l)  \cX^{k-1,l-1}_{n-1} - m (K_1-k) \cX^{k,l-1}_{n-1}+ n (K_2-l) \cX^{k-1,l}_n&= 0 \label{eq:Xiden2}
\end{align}
which are also a consequence of Yangian symmetry.

\section{Simplifying $\cY$}\label{simpY}

\subsection{Factorizing in the presence of Zhukowski variables} \label{secFactor}

With $u_i^\pm = u_i \pm \frac{iK_i}{g}$, we define\footnote{As in \cite{glebBound} we use the ``string scaling'' which differs from that of \cite{beiStau}.}
\beq
x^\pm + \frac{1}{x^\pm} \, = \, u^\pm \label{defX}
\eeq
and so
\beq
(x^\pm)^2 \, = \, x^\pm  \, u^\pm - 1 \, . \label{rule0}
\eeq
By way of example,
\beq
(x^+_1 - x^-_2)(x^+_1 - x^+_2) \, = \, (x^+_1)^2 - x^-_2 x^+_1 - x^+_1 x^+_2 + x^-_2 x^+_2 \, = \, (x^+_1 \, u_1^+ - 1) - x^-_2 x^+_1 - x^+_1 x^+_2 + x^-_2 x^+_2 \, . \label{example}
\eeq
To reverse this step is non-trivial: the expression on the r.h.s. of the last equation cannot be factored without knowledge of the square root property of the $x^\pm$ function defined by equation \eqref{defX}. In particular, algebraic computing systems are able to factor polynomials in variables like $g, \, u^\pm$ that do not obey such relations, but cannot easily be taught to apply rules like undoing \eqref{rule0}.

On the other hand (the two $\pm$ are independent),
\beq
(x^\pm_1 - x^\pm_2)  (1 - \frac{1}{x^\pm_1 \, x^\pm_2}  ) \, = \, u_1^\pm - u_2^\pm \, .\label{niklasRel}
\eeq
For a proof it suffices to expand the product  and to use \eqref{defX}. In a manner of speaking, the two factors on the left hand side are inverses of each other w.r.t. our factorisation issue, because the r.h.s. only contains variables that e.g. Factor[] in {\tt Mathematica} can handle. 

From \eqref{defX} we abstract the two replacement rules
\beq
(x^\pm)^{-n} \rar (x^\pm)^{-n+1}  (  u^\pm - x^\pm   ) \, , \qquad n \in \mathbb{N} \label{rule1}
\eeq
and
\beq
(x^\pm)^n \rar (x^\pm)^{n-2}  ( x^\pm \, u^\pm - 1   ) \, , \qquad n -1 \in \mathbb{N} \, . \label{rule2}
\eeq
We can use the property to simplify the r.h.s. of \eqref{example}: in a first step we multiply e.g. with the ``inverse'' of $(x^+_1-x^+_2)$
\beq
 (x^+_1 \, u^+ - 1 - x^-_2 x^+_1 - x^+_1 x^+_2 + x^-_2 x^+_2)  (1 - \frac{1}{x^+_1 \, x^+_2}  ) \, = \, (x^+_1 \, u^+ - 1 - x^-_2 x^+_1 - x^+_1 x^+_2 + x^-_2 x^+_2) \frac{x^+_1 x^+_2 - 1}{x^+_1 x^+_2}
\eeq
upon which we use \eqref{rule1} to eliminate $x^+_1, x^+_2$ from the denominator. Multiplying out one obtains up to cubic powers of $x^+_1, x^+_2$, on which now \eqref{rule2} is used repeatedly. 
We obtain
\beq
(x^+_1 - x^-_2)(u_1^+ - u_2^+) \,  \, = \, (x^+_1 - x^-_2) (x^+_1 - x^+_2)  (1 - \frac{1}{x^+_1 \, x^+_2}  )
\eeq
where the factorisation of the l.h.s. is easily achieved by Factor[] because the result is by construction \emph{multilinear} in $x^+_1,x^-_1,x^+_2,x^-_2$ ($x^-_1$ does not occur in this example.) Last, we have used \eqref{niklasRel} backwards to rewrite $u_1^+ - u_2^+$ in terms of $x^+_1,x^+_2$. Cancelling the last factor we have shown the factorisation of the rhs of \eqref{example} as desired. To arrive at the same conclusion one can alternatively use the ``inverse'' of $x^+_1-x^-_2$.

This procedure seems a little involved, but it gives a way of factoring out any of $x^\pm - y^\pm$ ($\pm$ is again independent in the two terms) or $1 - 1/(x^\pm \, y^\pm)$: to test for the presence of such a factor, one multiplies by its inverse and takes the steps described above. If Factor[] is able to pull out $u_1^\pm - u_2^\pm$ we have succeeded. One can also eliminate (positive or negative) powers of $x^\pm,y^\pm$ or factors like $x^+_1 - x^-_1, 1-1/(x^+_1 \, x^-_1)$.

Admittedly, the method only works by ``shooting'' in that we have to try the inverse of any particular factor to detect it. This is not much of an obstacle as long as an idea about the form of the result exists. As we shall see, it is possible to deal with more general polynomials of $x^\pm, y^\pm$ in the same way.

\paragraph{Computing $\cY$.} 
Our first application of the technique concerns the simplification of the $\cY$ matrix. From \eqref{defY} we see that it is defined by a matrix equation in which
the matrix $A$ has to be inverted. Employing Kramer's rule $A^{-1} \, = \, A^\# / \mathrm{Det}(A)$ we find that the entries of the adjoint matrix are polynomials of up to seventh (total) order in the representation parameters $\mathbf{r}$, but maximally cubic in each of them. The determinant in the denominator is
\beq
\mathrm{Det}(A) \, = \, -(K_1-n)(K_2 -m) \,p( u, K_1, K_2, \mathbf{r})
\eeq
with a polynomial $p \, = \,  (u_1-u_2)^2 \, p_{uu} + K_1^2 \, p_{11} + K_1 \, K_2 \, p_{12} + K_2^2 \, p_{22}$, where the $p_{ij}$ are maximally of overall order eight in the representation parameters. Remarkably, $p$ does not depend on $k,l,n$, a first hint that it might be factorisable in the way sketched above. 

Indeed after some rewritings and running our factorisation scheme on that form of $p$ we find
\beq
p \, = \, - g^2 \, (x^-_1-x^+_2)(x^+_1-x^-_2) ( 1 - \frac{1}{x^-_1 \, x^-_2}  ) (1 - \frac{1}{x^+_1 \, x^+_2}  ) \, . \label{aDetFin}
\eeq
The greatest worry has disappeared: the denominator of the $\cY$ matrix does not have a complicated dependence on the coupling constant, we only see the bricks of the Beisert $S$-matrix \cite{beisertPSU22}!

For the ensuing attempt on factoring $\cY$ it is perhaps not necessary but surely convenient to appeal to the contiguity equations \eqref{eq:Xkplus}-\eqref{eq:Xlmin} to reduce the r.h.s. of \eqref{defY} to a different basis of $\cX$-matrices with index shifts. The most concise formulae seem to arise choosing $\{\cX^{k,l}_n\, ,\cX^{k-1,l}_n\,,\cX^{k,l-1}_{n-1}\}$. 
 
Intriguingly, in all entries of $\cY$, the coefficients of $\{\cX^{k-1,l}_n, \, \cX^{k,l}_n\}$ both acquire the same $x^\pm, y^\pm$-dependent coefficient\footnote{For $\cY^I_I$ this was already noticed in \cite{fivePt}. The expressions given in (A.6) in that article motivate the present study.}, followed by different albeit simple rational functions of $ v, \, K_1, \, K_2, \, k, \, l, \, n$. We will state these in a form where the contiguity relations are used to reintroduce another instance of $\cX$ --- $\cX^{k,l-1}_{n-1}$ to be precise --- in order to eliminate $\delta u$ from the coefficients. These expressions are strikingly simple.

\subsection{Simplified scattering}
The Y-matrix can be split into two different parts under component-wise multiplication
\begin{align}
&\cY^{k,l}_n \, = \mathbb{Y} \star \tilde \cY^{k,l}_n,
&&  \rightarrow
&&\big( \cY^{k,l}_n\big)^I_J \, = \big(\mathbb{Y} \big)^I_J \big( \tilde \cY^{k,l}_n\big)^I_J.
\end{align}
The part $\mathbb{Y}$ depends only on the Zhukowski variables $x^\pm$. Recall that $U_i=\sqrt{x^+_i/x^-_i}$ and $\gamma_i$ is the representation parameter for the $i$-th particle.
\begin{align}
\mathbb{Y} = 
\begin{pmatrix}
\frac{x^+_1-x^+_2}{x^-_1-x^+_2} \frac{1}{U_1} &  \frac{x^+_2-x^-_2}{x^-_1-x^+_2} \frac{\gamma_1U_2} {\gamma_2U_1} & 0 & \frac{(x^+_1-x^-_1)(x^+_2-x^-_2)}{1-x^+_1x^+_2}\frac{iU_1U_2}{\gamma_1\gamma_2} \\
\frac{x^+_1-x^-_1}{x^-_1-x^+_2} \frac{\gamma_2} {\gamma_1} &  \frac{x^-_1-x^-_2}{x^-_1-x^+_2} \scriptstyle{U_2}  & \frac{(x^+_1-x^-_1)(x^+_2-x^-_2)}{x^-_1x^-_2-1}\frac{i}{\gamma_1\gamma_2} & 0 \\
0 & \frac{i}{1-x^-_1x^-_2}\frac{\gamma _1 \gamma _2}{U_1 U_2} & \frac{x^-_1-x^-_2}{x^+_1-x^-_2}\scriptstyle{U_1} &\frac{x^-_1-x^+_1}{x^+_1-x^-_2} \frac{\gamma _2 U_1}{\gamma _1 U_2}\\
\frac{i\gamma _1 \gamma _2}{x^+_1x^+_2-1} & 0 & \frac{x^-_2-x^+_2}{x^+_1-x^-_2}\frac{\gamma_1}{\gamma_2} & \frac{x^+_1 -x^+_2}{x^+_1-x^-_2}\frac{1}{U_2}
\end{pmatrix}
\end{align}
while $\tilde{\cY} = \tilde{\cY}_1+\tilde{\cY}_2+\tilde{\cY}_3$ only depends on $\cX, \delta v$ and simple numerical factors
\begin{align}
 \tilde{\cY}_1&= 
\begin{pmatrix}
 0 & 0 & 0 & 0 \\
 \frac{l}{\sqrt{K_1 K_2}} & \frac{l}{\delta v-\delta K} & \frac{1}{\sqrt{K_1 K_2}}
   & 0 \\
 0 & 0 & \frac{n-K_1}{\delta v-\delta K} & 0 \\
 \frac{l  (K_1-n  )}{\sqrt{K_1 K_2}} & 0 & \frac{K_1-n}{\sqrt{K_1 K_2}} & 0
\end{pmatrix} (\cX^{k,l-1}_{n-1}-\cX^{k,l}_n ), \\
 \tilde{\cY}_2&= 
\begin{pmatrix}
 \frac{k}{\delta v+\delta K} & \frac{k}{\sqrt{K_1 K_2}} & 0 & \frac{1}{\sqrt{K_1
   K_2}} \\
 0 & 0 & 0 & 0 \\
 0 & \frac{k  (K_2-m  )}{\sqrt{K_1 K_2}} & 0 & \frac{K_2-m}{\sqrt{K_1 K_2}} \\
 0 & 0 & 0 & \frac{m-K_2}{\delta v+\delta K}
\end{pmatrix} (\cX^{k-1,l}_n -\cX^{k,l}_n ),\\
 \tilde{\cY}_3&= 
\begin{pmatrix}
 1 & \sqrt{\frac{K_1}{K_2}} & 0 & 0 \\
 \sqrt{\frac{K_2}{K_1}} & 1 & 0 & 0 \\
 0 & \frac{k K_2-K_1 m}{\sqrt{K_1 K_2}} & 1 & \sqrt{\frac{K_2}{K_1}} \\
 \frac{K_1 l-K_2 n}{\sqrt{K_1 K_2}} & 0 & \sqrt{\frac{K_1}{K_2}} & 1 
\end{pmatrix} \cX^{k,l}_n .
\end{align}
Notice that this form makes the pole structure explicit; in particular, it has no spurious poles. At this point it is also easy to see the coefficients of the fundamental $S$-matrix appear since they simply correspond to the elements of $\mathbb{Y}$.

However, owing to the identity \eqref{eq:Xiden1} we can actually simplify the explicit $v$ dependence and write $\tilde{\cY}$ in the form of a compact matrix when $n \, \neq \, k$
\beq
\tilde \cY =\frac{1}{\sqrt{K_1K_2}} \begin{pmatrix}
\frac{\sqrt{K_1K_2}}{v+\delta K}\vecy{\frac{ l \, (n-K_1)}{k - n}}{\frac{k \, (K_2-l)}{k - n}}{l-K_2} &
\vecy{0}{k}{K_1-k} &
\vecy{0}{0}{0} &
\vecy{0}{1}{-1} \\
\vecy{l}{0}{K_2-l} & 
\frac{\sqrt{K_1K_2}}{v-\delta K}\vecy{\frac{ l \, (k-K_1)}{k - n}}{\frac{ k \, (K_2-m)}{k - n}}{ k-K_1} & 
\vecy{-1}{0}{1} &
\vecy{0}{0}{0} \\
\vecy{0}{0}{0} &
\vecy{0}{k \, (m-K_2)}{(K_1-k)m} &
\frac{\sqrt{K_1K_2}}{v-\delta K}\vecy{\frac{(n-K_1 )m}{k - n}}{\frac{k \, (K_2-m)}{k - n}}{m} & 
\vecy{0}{(m-K_2)}{-m} \\
\vecy{l \, (K_1 - n)}{0}{n \, (l-K_2)} &
\vecy{0}{0}{0} &
\vecy{(n-K_1 )}{0}{-n} &
\frac{\sqrt{K_1K_2}}{v+\delta K}\vecy{\frac{ l \, (n-K_1 )}{k - n}}{\frac{n \,  (K_2- m)}{k - n}}{n}
\end{pmatrix} \, . \label{yNonSing}
\eeq
The three-vectors refer to the ``basis'' $\{\cX^{k,l-1}_{n-1}, \, \cX^{k-1,l}_n, \, \cX^{k,l}_n\}$. Equation \eqref{yNonSing} can be taken as a definition, valid when $n \, \neq \, k$. 

\section{Simplifying $\cZ$} \label{simpZ}

\subsection{Factorization}

Similar to the derivation of $\cY$, in \cite{glebBound} the $\cZ$ matrix is found from a matrix equation \eqref{defZ} where $\cY'$ is a $6 \times 8$ block diagonal compilation of $\cY$ elements with index shifts $(k-1,l,n),(k,l-1,n),(k-1,l,n-1),(k,l-1,n-1)$ and the matrices $C, \, D$ depend on the representation parameters and the various counters. The inverse of $C$ needed to compute $\cZ$ is much simpler than that of $A$ discussed above.  However, all components of $(C)^{-1}$ have the denominator factor
\beq
d \, = \, x^-_1 \, x^-_2 - x^+_1 \, x^+_2 
\eeq
which can hardly be a physical singularity of the $S$-matrix; for once, in the residue calculation \cite{fivePt} the matrix elements are ``mirrror rotated'' $x^-_1 \rar 1/x^-_1, \, x^+_2 \rar 1/x^+_2$ and expanded to leading order in the coupling constant, so that $d$ yields a singularity
\beq
d' \, = \,  v_1^2 - v_2^2 - \frac{1}{4}(K_1^2 - K_2^2)
\eeq
which would spoil any hope of obtaining a Taylor series. In fact, upon explicitly evaluating the diagonal $\cZ$ elements in this kinematics and to leading order in $g$ it was seen in \cite{fivePt} that the singularity $d'$ generically cancels. Obviously one will ask whether the original denominator $d$ cancels from the full $\cZ$ matrix in the first place.

In order to apply the factorisation approach of Section \ref{secFactor} we need to construct a multiplicative ``inverse'' of $d$. To this end we write a general ansatz
\beq
e \, = \, \sum_{i,j,k,l \, = \, 0}^1 p_{ijkl}(u,v,K_1,K_2) \, (x^-_1)^i (x^-_2)^j (x^+_1)^k (x^+_2)^l \, , \label{defE}
\eeq
whose product with $d$ will also take the form
\beq
d \, e \, = \, \sum_{i,j,k,l \, = \, 0}^1 c_{ijkl}(u,v,K_1,K_2) \, (x^-_1)^i (x^-_2)^j (x^+_1)^k (x^+_2)^l
\eeq
upon employing \eqref{rule2}. Imposing $c_{ijkl} \, = \, 0 : i+j+k+l>0$ we obtain a set of 15 independent homogeneous equations on the 16 coefficients $p_{ijkl}$. Up to overall rescalings, the solution is unique. Choosing to scale up by the denominator we obtain the coefficients $p_{ijkl}$. With this scaling
\begin{eqnarray}
d \, e & = & c_{0000} \\ & = & (u^-)^4 - 2 \, (u^-)^2 (v^-)^2 + (v^-)^4 - 2 \, (u^-)^2 (u^+)^2 -
  2 \, (v^-)^2 (u^+)^2 + (u^-)^2 (v^-)^2 (u^+)^2 + (u^+)^4 \nonumber \\ && + 8 \, u^- v^- u^+ v^+ -
 (u^-)^3 v^- u^+ v^+ - u^- (v^-)^3 u^+ v^+ - u^- v^- (u^+)^3 v^+ - 2 \, (u^-)^2 (v^+)^2  - 2 \,
 (v^-)^2 (v^+)^2 \nonumber \\ &&+ (u^-)^2 (v^-)^2 (v^+)^2 - 2 \, (u^+)^2 (v^+)^2 + 
 (u^-)^2 (u^+)^2 (v^+)^2 + (v^-)^2 (u^+)^2 (v^+)^2 - u^- v^- u^+ (v^+)^3 + (v^+)^4 \nonumber
\end{eqnarray}
As for the simpler factorisation problems described above, if multiplying $e$ on any given polynomial and using the rule \eqref{rule2} yields a factor $c_{0000}$, we will have detected a factor $d$ in that polynomial. Finally, $c_{0000}$ can be cancelled against $d \, e$ in the denominator.

To not overcharge {\tt Mathematica}, it is helpful to decompose the test polynomial, say, $f$ in the same way as $e$ in \eqref{defE}. The product with $e$ is best taken keeping the coefficients in both polynomials abstract, leading to a decomposition of the type
$\sum \ldots p_\mathbf{i} \, q_\mathbf{j} \, = \, r_\mathbf{k}$ for the decomposition of the result in terms of the sixteen ``basis elements". The dots stand for coefficients expressed in terms of $u^\pm, v^\pm$. 

To start on simplifying $\cZ$ we reduce the problem to the calculation of two coefficient matrices for $\cX^{k-1,l}_n, \, \cX^{k,l}_n$ using the contiguity relations \eqref{eq:Xkplus}-\eqref{eq:Xlmin}. This is imperative here, only in such a form do all entries in the coefficient matrices factor out $c_{0000}$ upon multiplication by $e$. Barring for $\cZ^I_I, \, i \in \{1 \ldots 4\}$ and $\cZ^5_6, \, \cZ^6_5$ the computation is now as for $\cY$: in any other component, the sixteen $r_\mathbf{k}$ for $\cX^{k-1,l}_n$ have a common --- at times fairly involved --- polynomial factor depending on $v, K_1, K_2, k, l, n$, and the same happens for those multiplying $\cX^{k,l}_n$. These two ``long'' polynomials are in general distinct. The remaining simple factors and the powers $(x^-_1)^i (x^-_2)^j (x^+_1)^k (x^+_2)^l$ are finally put together and dealt with as sketched in Section \ref{secFactor} and its application to $\cY$. Like it happens for $\cY$ we obtain the same rational function of $x_{1,2}^\pm$ in the coefficients of both $\cX$'s. To illustrate 
these features we display the final expression for $\cZ^3_2$, which is the most concise example:
\begin{align}
\cZ^3_2  = & \frac{(x^-_1 - x^+_1)(x^-_2 - x^+_2)}{(x^-_1 - x^+_2)(x^+_1 - x^-_2)} \biggl( \frac{k \,[\delta \tilde u (k^2 + k \, l - m\,  K_1  - k \, K_2 + K_1 \, K_2 - k \, n + l \, n ) + l \, n \, (K_1+K_2)]}{K_1\, K_2 \, (\delta \tilde u + k + l) \, l} \, \cX^{k-1,l}_n \nonumber \\ 
& \quad + \frac{(K_1-k)[ \delta \tilde u^2 (n-k) + \delta \tilde u (k -n)\, (k+l-K_1 - K_2)   + l \, n \, (K_1 + K_2)]}{K_1\, K_2 \, (\delta \tilde u + k + l) \, l} \, \cX^{k,l}_n \biggr) \label{Z32}
\end{align}
In the last formula $\delta \tilde u \, = \, v_1^- - v_2^+$. The numerator factors in the square brackets are essentially what we called the ``long polynomials'' above. 

In the six special cases there are several different such polynomials within either set of sixteen $r$ coefficients. For $\cZ^5_6, \, \cZ^6_5$ one straightforwardly sees that there are minimally two $x^\pm, \, y^\pm$ structures: a trivial one producing an isolated instance of $\cX$, the other a problem similar to the simplification of $\cY$ and the more ordinary components of $\cZ$. Indeed, such formulae were pre-empted in \cite{fivePt}, equations (A.8), (A.9):
\beq
(\cZ^{k,l}_n)^5_6 \, = \, \cX^{k-1,l}_n - (\cZ^{k,l}_{n+1})^6_6 \, , \qquad (\cZ^{k,l}_n)^6_5 \, = \, \cX^{k,l-1}_{n-1} - (\cZ^{k,l}_{n-1})^5_5 \label{preEmpt}
\eeq
With some hindsight and a lot of patience we could find a similar split into two groups of terms also in the remaining four cases, where it is far less obvious how the long polynomials combine. Such a writing is, of course, not unique.

The coefficients of the $\cX$ matrices in \eqref{Z32} are generic in the following sense: the $K_1 \, K_2$ denominator occurs in all elements of the $\cZ^I_I, \, I \in \{1 \ldots 4 \}$ block and the fermionic blocks have $\eta_1 \eta_2 / (g \, \sqrt{K_1 \, K_2})$ while the $\{5,6\}$ block shows no such factor. Further, there is a simple pole $1/(\delta \tilde u + k + l)$ and perhaps some other simple denominator factors without $\delta u$. Last, for $\cZ^i_J, \, i \in \{1 \ldots 6 \}, \, J \in \{1 \ldots 4\}$ the two long polynomials are of order $O(\delta u), \, O(\delta u^2)$, respectively. For $\cZ^i_J, \, J \in \{5,6\}$ one finds $O(\delta u^2), \, O(\delta u^3)$ instead. Exceptions to the latter rule of thumb are only $\cZ^5_6$ and $\cZ^6_5$ whose numerators are of order $O(\delta u^3), \, O(\delta u)$ and $O(\delta u^3), \, O(\delta u^4)$, respectively. We will not elaborate on these two somewhat atypical cases in the following as they are given by $\cZ^5_5, \, \cZ^6_6$ through \eqref{preEmpt}.

Expressing $\cX^{k+\delta k,l+\delta l}_{n + \delta n}$ by the contiguous $\cX^{k-1,l}_n, \, \cX^{k,l}_n$ using \eqref{eq:Xkplus}-\eqref{eq:Xlmin} we obtain coefficients resembling those in \eqref{Z32}. Conversely, can the $\cY, \, \cZ$ elements be cast into a simpler form using more instances of $\cX$? Scanning the range $\delta k, \, \delta l, \, \delta n \in \{-2 \ldots 1\}$ it is found that some of the index shifts with $\delta n \, = \, \delta l$ are individually of the same form as \eqref{Z32}: there is one simple pole at $\delta \tilde u + k + l$ or no pole in $\delta u$, and the numerators of the two coefficients are of comparable order:
\begin{center}
\begin{tabular}{c | c | c}
$(\delta k, \delta l = \delta n)$ & $\{ O(\delta u^r), O(\delta u^s) \}$ & $1/(\delta \tilde u + k + l)$ \\
\hline
$(-2,0)$ & $\{ \delta u^2, \delta u\}$ & $1/(\delta \tilde u + k + l)$ \\
$(-1,-1)$ & $\{ 1, \delta u\}$ & $1/(\delta \tilde u + k + l)$ \\
$(0,-2)$ & $\{ \delta u^2, \delta u^3\}$ & $1/(\delta \tilde u + k + l)$ \\
\hline
$(-2,1)$ & $\{ \delta u^2, \delta u\}$ & $1$ \\
$(-1,1)$ & $\{ \delta u, 1\}$ & $1$ \\
$(0,-1)$ & $\{ 1, \delta u\}$ & $1$ \\
$(1,-2)$ & $\{ \delta u^2, \delta u^3\}$ & $1$ \\
$(1,-1)$ & $\{ \delta u, \delta u^2\}$ & $1$
\end{tabular}
\vskip 0.25 cm
Properties of the decomposition of $(\delta k, \delta l) \, = \, \cX^{k+\delta k, l+\delta l}_{n+\delta l}$ in terms of $\cX^{k-1,l}_n, \, \cX^{k,l}_n$
\end{center}
Other cases, especially when the range is extended to larger shifts, introduce new types of poles in $\delta u$. 

Attempting to use, say, \eqref{Z32} in an analytic resummation of residues as in \cite{fivePt} one would ideally want to construct a form in which each $\cX$ is multiplied by simple factors that can be absorbed into the defining $_4F_3$. Leaving this programme to future work, we propose here to eliminate $\delta u$ from the coefficients, which must already entail a simplification because a variable is suppressed. This is in fact possible as long as $n \, \neq \, k$: with the notation of the table above, we may use $(-1,-1)$ to subtract out the pole in $\delta u$, upon which also the order in $\delta u$ of the two long polynomials decreases by one unit. Successively, $(1,-1), \, (-1,1), \, (0,-1)$ can be employed to subtract powers of $\delta u$ from the higher to the lower orders. For instance,
\begin{eqnarray}
\cZ^3_2 & = & \frac{(x^-_1 - x^+_1)(x^-_2 - x^+_2)}{(x^-_1 - x^+_2)(x^+_1 - x^-_2)} \frac{1}{K_1 \, K_2} \, \Bigl( \, \bigl[l \, (K_2 - k-l+n) \bigr] \, (-1,1) + \bigl[l \, (k+l-n) \bigr] \, (0,-1) \nonumber \\
& & \qquad + \bigl[(K_2-l)(K_2 - k-l+n) \bigr] \, (-1,0) + \bigl[(K_2-l)(k+l-n)\bigr] \, (0,0)  \Bigr) \label{forInstance}
\end{eqnarray}
where we have written $(-1,0), \, (0,0)$ for $\cX^{k-1,l}_n, \, \cX^{k,l}_n$. In order to write $\cZ$ in terms of shifted $\cY$ elements it will prove useful to trade $(-1,1), \, (1,-1)$ for $(-2,0), (0,-2)$ by the five-term identity
\begin{eqnarray}
0 & = & \phantom{-} \bigl[ k \, (K_2 - k - l + n) \bigr] \, (-1,0) - 
\bigl[ l \, (K_1 - n) \bigr] \, (0,-1)  - 
\bigl[ (k-n) (K_1 + K_2 - 2 \, k - 2 \, l  - 2) \bigr] \, (0,0) \nonumber \\ && -
\bigl[ (n+1) (K_2 - l - 1) \bigr] \, (0,1) + 
\bigl[ (K_1 - k - 1) (k + l - n + 1) \bigr] \, (1,0) \, .
\end{eqnarray}

\subsection{$\cZ$ from $\cY$}

After the appropriate simplifications, we found a very compact and interesting way to define the $\cZ$ block. It can be expressed quadratically in the $\cY$ block by introducing a wedge product so that we  can write $\cZ = \cY \wedge \cY$. On the level of the basis vectors we identify
\begin{align}
|k,l\rangle^{(0)}_1 &\simeq |k,l\rangle^{(0)}_1 \wedge |k,l\rangle^{(0)}_2\\
|k,l\rangle^{(0)}_2 &\simeq |k,l\rangle^{(0)}_3 \wedge |k,l\rangle^{(0)}_2\\
|k,l\rangle^{(0)}_3 &\simeq |k,l\rangle^{(0)}_1 \wedge |k,l\rangle^{(0)}_4\\
|k,l\rangle^{(0)}_4 &\simeq |k,l\rangle^{(0)}_3 \wedge |k,l\rangle^{(0)}_4\\
|k,l\rangle^{(0)}_5 &\simeq |k,l\rangle^{(0)}_3 \wedge |k,l\rangle^{(0)}_1\\
|k,l\rangle^{(0)}_6 &\simeq |k,l\rangle^{(0)}_4 \wedge |k,l\rangle^{(0)}_2
\end{align}
Then
\begin{align}
&(\cZ )^I_J= (\cY)^a_c \wedge (\cY)^b_d =  (\cY)^a_c \cdot  (\cY)^b_d - (\cY)^a_d \cdot  (\cY)^b_c  ,
\end{align}
with $ |k,l\rangle^{(0)}_I \simeq |k,l\rangle^{(0)}_a \wedge |k,l\rangle^{(0)}_b,
|k,l\rangle^{(0)}_J \simeq |k,l\rangle^{(0)}_c \wedge |k,l\rangle^{(0)}_d$ and the product acts on $\cX$ as
\begin{align}
&\cX^{kl}_n\cdot \cX^{kl}_n = \cX^{kl}_n,
&&\cX^{kl}_n\cdot \cX^{k+a,l+b}_{n+b} = \cX^{k+a,l+b}_{n+b},
&&\cX^{k-1,l}_n\cdot \cX^{k,l-1}_{n-1} = \cX^{k-1,l-1}_{n-1}.
\end{align}
Using \eqref{eq:Xiden1} we always make sure that one of the $\cY$ factors has a $\cX^{k-1,l}_n$ term and the other has a term $\cX^{k,l-1}_{n-1}$. This ensures that any component of $\cZ$ can be written as a linear combination of $\cX^{kl}_n,\cX^{k-1,l}_n,\cX^{k,l-1}_{n-1},\cX^{k-1,l-1}_{n-1}$.  Because of the identities that $\cX$ satisfies, it does not matter which $\cY$ factor has the $\cX^{k-1,l}_n$ term. From this we find the additional rules
\begin{align}
&\cX^{k-1,l}_n\cdot \cX^{k-1,l}_{n } = \frac{l (K_1-n)}{k(K_2-m)}\cX^{k-1,l-1}_{n-1} +  \frac{(k-n)(v+\sK-k-l)}{k(K_2-m)}\cX^{k-1,l}_n\\
&\cX^{k,l-1}_{n-1}\cdot \cX^{k,l-1}_{n-1} = \frac{k (K_2-m)}{l(K_1-n)}\cX^{k-1,l-1}_{n-1} +  \frac{(k-n)(v+\sK-k-l)}{l(K_1-n)}\cX^{k,l-1}_{n-1}.
\end{align}
Recall that we can write $\tilde{\cY}$ as a three-vector w.r.t. the spanning system $\{\cX^{k,l-1}_{n-1}, \, \cX^{k-1,l}_n, \, \cX^{k,l}_n\}$. In particular, let $(\cY)^a_b = \vecy{0}{y_1}{z_1}$ and $(\cY)^c_d = \vecy{x_2}{0}{z_2}$. Then from the above rules we find the very compact expression
\begin{align}\label{doubleYBracket}
\vecy{0}{y_1}{z_1} \cdot \vecy{x_2}{0}{z_2} \equiv  
y_1 \big(\tilde{\cY}^{k-1,l}_{n}\big)^c_d + z_1 \big(\tilde{\cY}^{k,l}_{n}\big)^c_d =: 
\vecy{0}{y_1}{z_1}^c_d \stackrel{!}{=} \vecy{x_2}{0}{z_2}^a_b. 
\end{align}
The new symbol $\bigl[\bigr]^c_d$ denotes a decomposition in terms of $\{\cY^{k,l-1,c}_{n-1,d}, \, \cY^{k-1,l,c}_{n,d}, \, \cY^{k,l,c}_{n,d}\}$ as is apparent from the middle part of the last equation. The equality at the very right is a non-trivial consequence of the form of $\tilde{\cY}$ and the properties of $\cX$; so there are always two equivalent ways of decomposing in terms of $\cY$-elements with index shifts. 

This seems to be a type of fusion relation in which the scattering of two bosons is written as some sort of composite scattering of fermions. At this point it is unclear what the meaning of this observation is, but it hints at some further structure of the bound state S-matrix. Understanding this property might be important, for example, for potentially finding a universal $R$-matrix. It would be interesting to understand the nature of the wedge product and its non-trivial action on $\cX$.

As an example, let us work out $(\cZ)^1_2$. Via the above identification, we have $|k,l\rangle^{(0)}_1 \simeq |k,l\rangle^{(0)}_1 \wedge |k,l\rangle^{(0)}_2$ and $|k,l\rangle^{(0)}_2 \simeq |k,l\rangle^{(0)}_3 \wedge |k,l\rangle^{(0)}_2$. Thus
\begin{align}
(\cZ)^1_2 &= (\cY)^1_3 \cdot (\cY)^2_1-(\cY)^1_2 \cdot (\cY)^2_3 \\
&=  \frac{x^+_2-x^-_2}{x^-_1-x^+_2} \frac{\gamma_1U_2} {\gamma_2U_1} \cdot\frac{(x^+_1-x^-_1)(x^+_2-x^-_2)}{1-x^-_1x^-_2}\frac{1}{\gamma_1 \gamma_2} 
\Big[\frac{k }{\sqrt{K_1K_2}}\cX^{k-1,l}_n + \frac{K_1-k}{\sqrt{K_1K_2}}\cX^{kl}_n\Big]\cdot\Big[\frac{\cX^{k,l-1}_{n-1}-\cX^{kl}_n}{\sqrt{K_1 K_2}}\Big]\\
&=   \frac{(x^+_1-x^-_1)(x^+_2-x^-_2)^2}{(x^-_1-x^+_2)(1-x^-_1x^-_2)}\frac{U_2}{U_1 \gamma^2_2}  \Big[\frac{k}{K_1K_2} (\cX^{k-1,l-1}_{n-1}- \cX^{k-1,l}_{n})+\frac{K_1-k}{K_1K_2}(\cX^{k,l-1}_{n-1}-\cX^{kl}_n)\Big] \\
& =  \frac{(x^+_1-x^-_1)(x^+_2-x^-_2)^2}{(x^-_1-x^+_2)(1-x^-_1x^-_2)}\frac{U_2}{U_1 \gamma^2_2} \frac{1}{\sqrt{K_1K_2}} \vecy{-1}{0}{1}^1_2. 
\end{align}
Since $\cY^1_3=\cY^3_1=\cY^2_4=\cY^4_2=0$, we see that almost all components of $\cZ$ are just given by one term. However, this is not true for the diagonal elements $\cZ^i_i$, where $i=1,2,3,4$ and $\cZ^5_6,\cZ^6_5$. As a consequence, these elements have two different $x^\pm$ dependent prefactors.

\subsection{Results for $\mathbf{\cZ}$} \label{resZ}

Following the decomposition of the wedge product, we can write
\begin{align}
\cZ = \mathbb{Z}_1 \star \tilde{Z}_1 -  \mathbb{Z}_2 \star \tilde{Z}_2,
\end{align}
where
\begin{align}
\mathbb{Z}_1 &= \begin{pmatrix}
 \mathbb{Y}_1^1 \mathbb{Y}_2^2 & 0 & 0 & 0 & 0 & \mathbb{Y}_2^2 \mathbb{Y}_4^1 \\
 0 & \mathbb{Y}_2^2 \mathbb{Y}_3^3 & 0 & 0 & \mathbb{Y}_1^2 \mathbb{Y}_3^3 & \mathbb{Y}_2^2 \mathbb{Y}_4^3 \\
 0 & 0 & \mathbb{Y}_1^1 \mathbb{Y}_4^4 & 0 & 0 & 0 \\
 0 &0 & 0 & \mathbb{Y}_3^3 \mathbb{Y}_4^4 & \mathbb{Y}_1^4 \mathbb{Y}_3^3 & 0 \\
 0 & \mathbb{Y}_2^1 \mathbb{Y}_3^3 & 0 & \mathbb{Y}_3^3 \mathbb{Y}_4^1 & \mathbb{Y}_1^1 \mathbb{Y}_3^3 & \mathbb{Y}_2^1 \mathbb{Y}_4^3 \\
 \mathbb{Y}_1^4 \mathbb{Y}_2^2 & \mathbb{Y}_2^2 \mathbb{Y}_3^4 & 0 & 0 & \mathbb{Y}_1^2 \mathbb{Y}_3^4 & \mathbb{Y}_2^2 \mathbb{Y}_4^4 
\end{pmatrix}
&&\mathbb{Z}_2 = \begin{pmatrix}
 \mathbb{Y}_1^2 \mathbb{Y}_2^1 & \mathbb{Y}_2^1 \mathbb{Y}_3^2 & \mathbb{Y}_1^2 \mathbb{Y}_4^1 & \mathbb{Y}_3^2 \mathbb{Y}_4^1& \mathbb{Y}_1^1 \mathbb{Y}_3^2 & 0 \\
\mathbb{Y}_2^3 \mathbb{Y}_1^2 & \mathbb{Y}_2^3 \mathbb{Y}_3^2 & \mathbb{Y}_1^2 \mathbb{Y}_4^3 & \mathbb{Y}_3^2 \mathbb{Y}_4^3 & 0 & 0 \\
 \mathbb{Y}_1^4 \mathbb{Y}_2^1 & \mathbb{Y}_2^1 \mathbb{Y}_3^4 & \mathbb{Y}_1^4 \mathbb{Y}_4^1 & \mathbb{Y}_3^4 \mathbb{Y}_4^1 & \mathbb{Y}_1^1 \mathbb{Y}_3^4 & \mathbb{Y}_2^1 \mathbb{Y}_4^4 \\
 \mathbb{Y}_1^4 \mathbb{Y}_2^3 & \mathbb{Y}_2^3 \mathbb{Y}_3^4 & \mathbb{Y}_1^4 \mathbb{Y}_4^3 & \mathbb{Y}_3^4 \mathbb{Y}_4^3 & 0 & \mathbb{Y}_2^3 \mathbb{Y}_4^4 \\
 \mathbb{Y}_1^1 \mathbb{Y}_2^3 & 0 & \mathbb{Y}_1^1 \mathbb{Y}_4^3 & 0 & 0 & \mathbb{Y}_2^3 \mathbb{Y}_4^1 \\
 0 & 0 & \mathbb{Y}_1^2 \mathbb{Y}_4^4 & \mathbb{Y}_3^2 \mathbb{Y}_4^4 & \mathbb{Y}_1^4 \mathbb{Y}_3^2 & 0
\end{pmatrix}
\end{align}
and
\begin{align}
\tilde{\cZ}_1 = \begin{pmatrix}
\vecy{\frac{l}{v^-_0}}{0}{\frac{v^-_l}{v^-_0}}^1_1 & 0 & 0 & 0 & 0 &\vecy{0}{\frac{1}{\sqrt{K_1K_2}}}{\frac{-1}{\sqrt{K_1K_2}}}^2_2 \\
 0 & \vecy{0}{\frac{k(K_2-m)}{(K_1-n)v^-_0}}{\frac{(K_1-k)v^-_m}{(K_1-n)v^-_0}}^3_3  & 0
   & 0 & \vecy{0}{\frac{k(m-K_2)}{l(v^-_0)}}{\frac{mv^+_{m-K_2}}{lv^-_0}}^2_1 &
\vecy{0}{\frac{K_2-m}{\sqrt{K_1K_2}}}{\frac{m}{\sqrt{K_1K_2}}}^2_2\\
 0 & 0 &\vecy{\frac{l(K_1-n)}{(K_2-m)v^+_0}}{0}{\frac{(K_2-l)v^+_n}{(K_2-m)v^+_0}}^4_4  & 0
   & 0 & 0 \\
 0 & 0 & 0 & \vecy{0}{\frac{m-K_2}{v^+_0}}{\frac{v^+_{m-K_2}}{v^+_0}}^3_3  &
 \vecy{0}{\frac{k(m-K_2)}{l v^-_0}}{\frac{m v^+_{k-K_2}}{lv^-_0}}^4_1 & 0 \\
 0 & \vecy{0}{\frac{k}{\sqrt{K_1K_2}}}{\frac{K_1-k}{\sqrt{K_1K_2}}}^3_2 & 0 &
\vecy{0}{\frac{1}{\sqrt{K_1 K_2}}}{\frac{-1}{\sqrt{K_1 K_2}}}^3_3  & 
\vecy{0}{\frac{k}{v^+_0}}{\frac{v^+_k}{v^+_0}}^3_3& 
\vecy{\frac{l(K_1-n)}{k\sqrt{K_1K_2}}}{0}{\frac{ln-(l-m)v^+_{-K_2}}{k\sqrt{K_1K_2}}}^1_2\\
\vecy{0}{\frac{k(K_2-m)}{(K_1-n)v^-_0}}{\frac{(K_1-k)v^-_m}{(K_1-n)v^-_0}}^4_1 &
\vecy{0}{\frac{k(K_2-m)}{(K_1-m)v^-_0}}{\frac{(K_1-k)v^-_m}{(K_1-m)v^-_0}}^4_3 & 0
   & 0 & \vecy{0}{\frac{k(K_2-m)}{l\sqrt{K_1K_2}}}{\frac{km-(k-n)v^+_{-K_2}}{l\sqrt{K_1K_2}}}^2_1 &
   \vecy{0}{\frac{m-K_2}{v^+_0}}{\frac{v^+_{m-K_2}}{v^+_0}}^2_2
\end{pmatrix}
\end{align}
\begin{align}
\tilde{\cZ}_2 =\begin{pmatrix}
 \vecy{0}{\frac{k}{\sqrt{K_1K_2}}}{\frac{K_1-k}{\sqrt{K_1K_2}}}^2_1 &
 \vecy{\frac{1}{\sqrt{K_1K_2}}}{0}{\frac{-1}{\sqrt{K_1K_2}}}^1_2 &
 \vecy{0}{\frac{1}{\sqrt{K_1K_2}}}{\frac{-1}{\sqrt{K_1K_2}}}^2_1 &
  \vecy{\frac{1}{\sqrt{K_1K_2}}}{0}{\frac{-1}{\sqrt{K_1K_2}}}^1_4&
  \vecy{\frac{1}{\sqrt{K_1K_2}}}{0}{\frac{-1}{\sqrt{K_1K_2}}}^1_1 & 
  0 
  \\
  \vecy{\frac{l}{\sqrt{K_1K_2}}}{0}{\frac{K_2-l}{\sqrt{K_1K_2}}}^3_2 &
   \vecy{\frac{1}{\sqrt{K_1K_2}}}{0}{\frac{-1}{\sqrt{K_1K_2}}}^3_2  &
    \vecy{\frac{l}{\sqrt{K_1K_2}}}{0}{\frac{K_2-l}{\sqrt{K_1K_2}}}^3_4 &
  \vecy{\frac{1}{\sqrt{K_1K_2}}}{0}{\frac{-1}{\sqrt{K_1K_2}}}^3_4  & 0 & 0
   \\
      \vecy{0}{\frac{k}{\sqrt{K_1K_2}}}{\frac{K_1-k}{\sqrt{K_1K_2}}}^4_1 & 
      \vecy{0}{\frac{k}{\sqrt{K_1K_2}}}{\frac{K_1-k}{\sqrt{K_1K_2}}}^4_3 & 
      \vecy{0}{\frac{1}{\sqrt{K_1K_2}}}{\frac{-1}{\sqrt{K_1K_2}}}^4_1 &
           \vecy{0}{\frac{1}{\sqrt{K_1K_2}}}{\frac{-1}{\sqrt{K_1K_2}}}^4_3 &
             \vecy{\frac{K_1-n}{\sqrt{K_1K_2}}}{0}{\frac{n}{\sqrt{K_1K_2}}}^1_1 & 
  \vecy{\frac{l(n-K_1)}{k \sqrt{K_1 K_2} (m-K_j)}}{0}{\frac{(n-k)(v^+_{l-K_2})+k}{k \sqrt{K_1 K_2} (m-K_j)}}^1_2
    \\
	 \vecy{0}{\frac{k(K_2-m)}{\sqrt{K_1K_2}}}{\frac{m(k-K_1)}{\sqrt{K_1K_2}}}^4_1&
	\vecy{\frac{K_1-n}{\sqrt{K_1K_2}}}{0}{\frac{n}{\sqrt{K_1K_2}}}^3_2  & 
	\vecy{0}{\frac{K_2-m}{\sqrt{K_1K_2}}}{\frac{m}{\sqrt{K_1K_2}}}^4_1 &
	\vecy{0}{\frac{K_2-m}{\sqrt{K_1K_2}}}{\frac{m}{\sqrt{K_1K_2}}}^4_3 & 0 
	&
	\vecy{\frac{l(n-K_1)}{k v^+_0}}{0}{\frac{nv^+_{l-K_2}}{kv^+_0}}^3_2
   \\
 \vecy{\frac{l(K_1-n)}{v^+_0}}{0}{\frac{(K_2-l)v^+_n}{v^+_0}}^3_2   & 
 0 &
  \vecy{\frac{l(K_1-n)}{v^+_0}}{0}{\frac{(K_2-l)v^+_n}{v^+_0}}^3_4  & 
  0 & 
  0 & 
  \vecy{\frac{l(n-K_1)}{k \sqrt{K_1 K_2} (m-K_j)}}{0}{\frac{(n-k)(v^+_{l-K_2})+k}{k \sqrt{K_1 K_2} (m-K_j)}}^3_2
    \\
 0 & 
 0 & 
 \vecy{\frac{l}{\sqrt{K_1K_2}}}{0}{\frac{K_2-l}{\sqrt{K_1K_2}}}^4_4&
 \vecy{\frac{1}{\sqrt{K_1K_2}}}{0}{\frac{-1}{\sqrt{K_1K_2}}}^4_4 &
 \vecy{0}{\frac{k(K_2-m)}{\sqrt{K_1K_2}}}{\frac{(n-k)(v^+_k-k(K_2-l))}{\sqrt{K_1K_2}}}^2_3 & 
 0 
\end{pmatrix}
\end{align}
For conciseness, we have defined
\begin{equation}
v^\pm = v \pm \frac{K}{2}, \qquad
v_a^{\pm} = v^{\pm}_1 - v^{\pm}_2 - a.
\end{equation}
As we can see, $\cZ^5_6$ and $\cZ^6_5$ cannot be very elegantly expressed in terms of $\cX^{k+\delta k,l+\delta l}_{n+\delta l}$. However, if we allow for atypical index shifts then they simplify, too, since from $\alg{su}(2)$ invariance we can prove
\begin{align}
&(\cZ^{k,l}_{n+1})^i_6 = - (\cZ^{k,l}_{n})^i_5,
&&(\cZ^{k+1,l-1}_{n})^6_i = - (\cZ^{k,l}_{n})^5_i, 
&&(\cZ^{k,l}_{n})^6_5 = (\cZ^{k+1,l-1}_{n-1})^5_6 .
\end{align}
and
\begin{align}
& (\cZ^{k,l}_{n})^5_5 &= \cX^{k,l-1}_n - (\cZ^{k,l}_{n+1})^6_5, 
 &&(\cZ^{k,l}_{n})^5_6 &= \cX^{k-1,l}_n - (\cZ^{k,l}_{n+1})^6_6,
 &&(\cZ^{k,l}_{n})^5_5 &= (\cZ^{k+1,l-1}_{n+1})^6_6 .
\end{align}
We have checked that these relations indeed hold.

We would like to stress again that the decomposition in terms of $\cX$ functions is not unique, once instance of \eqref{doubleYBracket}) is
\begin{align}
\vecy{0}{\frac{k(K_2-m)}{(K_1-n)v^-_0}}{\frac{(K_1-k)v^-_m}{(K_1-n)v^-_0}}^4_1 = 
\vecy{-l \, (K_1 - n+1)}{0}{n \, (K_2 - l)}^2_2.
\end{align}
relevant to the bottom left corner of $\tilde Z_1$. Consequently, there are also several ways to express $\cZ$ in terms of $\cY$.

\section{Properties}

In this section we discuss some properties of the bound state $S$-matrix. We will mainly generalize the properties that were found for the fundamental $S$-matrix, along the lines as they were formulated in \cite{Beisert:2014hya}.

\paragraph{Braiding and physical unitarity.}
Much like the $S$-matrix of fundamental particle, the bound-state $S$-matrix enjoys braiding unitarity,
\begin{align}
S_{12}(u_1,u_2) S_{21}(u_2,u_1) =1\,.
\end{align}
This provides us with a simple way to compute the inverse $S$-matrix, which is important when describing the scattering of particles in the anti-symmetric representation.

\paragraph{Generalised physical unitarity.}
If we started from a unitary representation of the symmetry algebra, \textit{e.g.}\ by picking $\gamma$ like in \eqref{eq:gamma}, the S-matrix also enjoys generalised physical unitarity
\begin{align}
S_{12}(u_1^*,u_2^*)^\dagger S_{12}(u_1,u_2) =1.
\end{align}
Complex conjugation acts on the  $S$-matrix parameters as
\begin{align}
(x^\pm_k,\gamma_k,v_k)^* = (x_k^\mp,\frac{i}{U_k}\gamma_k,-v_k)\,,
\end{align}
where $\gamma$ is given by~\eqref{eq:gamma}.

\paragraph{Symmetry.}
For $\gamma$ like in \eqref{eq:gamma}, we find that the S-matrix is symmetric:
\begin{align}
\frac{S^A_B}{\langle A| A\rangle} = \frac{S^B_A}{\langle B| B\rangle}.
\end{align}
This property is easy to prove from \eqref{eq:SymX} and the explicit form of $\cY$ and $\cZ$ in terms of $\cX$. Thus, if we properly normalize our states, then this reduces to the regular relation $S^T=S$.
\paragraph{Inversion.}
By combining the symmetry property and physical unitarity we find that the inverse $S$-matrix may be computed by sending 
\begin{align}
(x^\pm_k,\gamma_k,v_k)  \rightarrow (x_k^\mp,\frac{i}{U_k}\gamma_k,-v_k)\,. 
\end{align}
Remarkably, this property holds for any~$\gamma$.
\paragraph{Crossing.}
It is most convenient to define crossing symmetry analogous to \cite{Beisert:2014hya}. The charge conjugation transformation then simply corresponds to the trivial automorphism
\begin{align}
\mathcal{C}\cdot |a,b,c,d\rangle = i^{a+b+c+d} (-1)^{a+c} |b,a,d,c\rangle.
\end{align}
The prefactor $ i^{a+b+c+d}  = i^K$ is used for convenience.
It corresponds to the simple transformation that acts on the variables that generate the bound state representation as
\begin{equation}
w_i \mapsto \epsilon^{ij} w_j \, , \qquad
\theta_\alpha  \mapsto  \epsilon^{\alpha\beta} \theta_\beta.
\end{equation}
From this it is easy to see that
\begin{equation}
\mathcal{C}^2=1 \, , \qquad [\mathcal{C}\otimes \mathcal{C}]\, S\, [\mathcal{C}^{-1}\otimes \mathcal{C}^{-1} ]= S.
\end{equation}
We the find the following crossing symmetry of the $S$-matrix, written in components as
\begin{align}
\Big(S^{-1}(u_1,u_2)\Big)^{|A\rangle \otimes |B\rangle}_{|C\rangle \otimes |D\rangle } = 
 \frac{\langle A | A \rangle}{\langle C | C \rangle}(-1)^{|C|(|A|+1)}F\Big(S(u_1^{cross},u_2)\Big)^{\mathcal{C}(|C\rangle)  \otimes |B\rangle }_{\mathcal{C}(|A\rangle) \otimes |D\rangle  }
\end{align}
where \cite{Arutyunov:2009kf}
\begin{align}
F = 
\frac{x^+_1-x^-_2}{x^-_1-x^-_2}\frac{\frac{1}{x^+_1}-x^+_2}{\frac{1}{x^-_1} - x^+_2} 
\prod_{\alpha =1}^{K_1-1} \frac{v+\dK -\alpha}{v - \dK+\alpha}.
\end{align}
and the crossing transformation is
\begin{align}
(x^\pm,\gamma,U) \rightarrow \Big(\frac{1}{x^\pm},i\frac{U-U^{-1}}{\gamma},\frac{1}{U}\Big)
\end{align}
Upon properly normalizing our basis elements, the crossing relation can now brought to the standard form
\begin{align}
(\mathcal{C}\otimes 1)\,S^{t_1}(u_1^{cross},u_2)\,(\mathcal{C}^{-1}\otimes 1) = FS^{-1}.
\end{align}
\paragraph{Monodromy.}
We have that $S$ is also invariant under the crossing (in the same way) both variables, \textit{i.e.}
\begin{align}
(x^\pm_1,\gamma_1;\ x^\pm_2,\gamma_2)   \rightarrow \Big(\frac{1}{x_1^\pm},\frac{i\gamma_1}{x^+_1},\ \frac{1}{x_2^\pm},\frac{i\gamma_2}{x^2_k}\Big) .
\end{align}
For a particular choice of $\gamma_i = \sqrt{i(x^+-x^-)U}$, see \textit{e.g.}~\cite{Arutyunov:2009ga}, this is precisely the crossing transformation. More generally, this corresponds to crossing transformation on $x^\pm$ combined with a redefinition of $\gamma$ which follows from a local basis transformation.

\section{Conclusions}

The construction of the bound state $S$-matrix in \cite{glebBound} is complete, though not completely explicit: one is left to work with certain matrix inverses which obfuscate for instance the pole structure. The central obstruction to simplification are the Zhukowsky variables $x^\pm, \, y^\pm$ that are root functions, which impede factorisation if occurring in rational functions. For the case at hand we solved this problem introducing a concept of ``inverse'' (modulo readily factorisable expressions) for certain combinations of Zhukowsky variables.

Our results are split into a part containing Zhukowsky variables, and with them the dependence of the bound state scattering matrix on the 't Hooft coupling $\lambda$, and another one of hypergeometric type. The first factor is of the same type as in the Beisert $S$-matrix for fundamental particles \cite{beisertPSU22}. It has only physical singularities, e.g. poles like $u^+ - v^-$ or $u^\pm$; for once, the unphysical $x^-_1 x^-_2 - x^+_1 x^+_2$ singularity of the $\cZ$ block is shown to cancel. 

The hypergeometric parts depend on the various counters and the rapidity difference, but not on $\lambda$. Its $\cY$ blocks can be expressed by $\cX$ elements with shifted counters, likewise those of $\cZ$ are written in terms of $\cY$; from where one can regain a slightly more complicated form in terms of $\cX$. We display completely explicit results for all parts on just a few pages. There are only a few distinct coefficients in these formulae; their appearance suggests that there may be a unifying superspace form. In particular, we have found a very suggestive relation between the $\cY$ and $\cZ$ components that hints at a fused structure.

Finally we have clarified several properties of the bound state $S$-matrix such as crossing, inversion and braiding unitarity. 

The writing we chose was mainly motivated by brevity; it is, of course, not unique. An open question is what form will be most useful for residue calculations as in \cite{shotaThiago2,fivePt} or alternative future approaches to multiple glueings of hexagon tiles. Our findings might also yield interesting reformulations of the TBA \cite{TBA}.

\paragraph{Acknowledgements}

MdL was supported by SFI, the Royal Society and the EPSRC for funding under grants UF160578, RGF$\backslash$EA$\backslash$181011, RGF$\backslash$EA$\backslash$180167 and 18/EPSRC/3590. AS's work is funded by ETH Career Seed Grant No. SEED-2319-1. BE and AS are supported by the Spark grant  n.~190657 ``Exact correlation functions in AdS/CFT'', as well as by the NCCR SwissMAP, funded by the Swiss National Science Foundation.

\appendix

\section{Notebook}

We have appended a \textit{Mathematica} notebook with all the components and relations of the $S$-matrix that are presented here. In this appendix we will briefly explain the notation of the notebook.

\begin{itemize}
    \item The components of the $S$-matrix are called 
\begin{align}
(\cX_{ij})^{kl}_n \ \leftrightarrow \ \mathtt{SmatX}[i,j][k,l,n]    \\
(\cY_{ij})^{kl}_n \ \leftrightarrow \ \mathtt{SmatY}[i,j][k,l,n]    \\
(\cZ_{ij})^{kl}_n \ \leftrightarrow \ \mathtt{SmatZ}[i,j][k,l,n]    
\end{align}    
    \item The basis vectors of the bound state representation are denoted by
\begin{align}
|a,b,c,d\rangle  \ \leftrightarrow \ \mathtt{state}[a,b,c,d]
\end{align} 
\item States can be multiplied using $\mathtt{CenterDot}$
\begin{align}
|a_1,b_1,c_1,d_1\rangle \otimes |a_2,b_2,c_2,d_2\rangle \ \leftrightarrow \ \mathtt{state}[a_1,b_1,c_1,d_1]\cdot\mathtt{state}[a_2,b_2,c_2,d_]
\end{align} 
\item The $S$-matrix is then programmed as an operator acting on such states as
\begin{align}
S|a_1,b_1,c_1,d_1\rangle \otimes |a_2,b_2,c_2,d_2\rangle \ \leftrightarrow \ \mathtt{S}[1,2][\mathtt{state}[a_1,b_1,c_1,d_1]\cdot\mathtt{state}[a_2,b_2,c_2,d_]],
\end{align}
which evaluates to give the correct components.
\item In order to not deal with spurious poles in $\cX,\cY,\cZ$, we send $K\mapsto K+\epsilon$ and send $\epsilon \rightarrow0$ in the end. This regulates combinatorial factors of the form $K_i-A$ which sometimes naively result in a $0/0$.
\end{itemize}


\begin{thebibliography}{999}

\bibitem{123}
J.~Maldacena, Adv.~Theor.~Math.~Phys. {\bf 2} (1998) 231 [hep-th/9711200];
S.~Gubser, I.~Klebanov and A.~Polyakov, Phys.~Lett. {\bf B428} (1998) 105 [hep-th/9802109];
E.~Witten, Adv.~Theor.~Math.~Phys. {\bf 2} (1998) 253 [hep-th/9802150].

\bibitem{beiStau}
D.~E.~Berenstein, J.~M.~Maldacena and H.~S.~Nastase,
JHEP {\bf 0204} (2002) 013 [hep-th/0202021];
J.~Minahan and K.~Zarembo, JHEP {\bf 0303} (2003) 013 [hep-th/0212208];
I.~Bena, J.~Polchinski and R.~Roiban, Phys.\ Rev. {\bf D69} (2004) 046002 [hep-th/0305116];
N.~Beisert, V.~Dippel and M.~Staudacher, JHEP {\bf 0407} (2004) 075 [hep-th/0405001];
N.~Beisert and M.~Staudacher, Nucl.~Phys. {\bf B727} (2005) 1 [hep-th/0504190].

\bibitem{beisertPSU22}
N.~Beisert,
Adv.\ Theor.\ Math.\ Phys.\  {\bf 12} (2008) 945 [hep-th/0511082].

\bibitem{BES}
N.~Beisert, B.~Eden and M.~Staudacher,
J.\ Stat.\ Mech.\  {\bf 0701} (2007) P01021 [hep-th/0610251].

\bibitem{luscher}
M.~Luscher,
Commun. Math. Phys. \textbf{104} (1986), 177
doi:10.1007/BF01211589;
M.~Luscher,
Commun. Math. Phys. \textbf{105} (1986), 153-188
doi:10.1007/BF01211097;
J.~Ambjorn, R.~A.~Janik and C.~Kristjansen,
Nucl. Phys. B \textbf{736} (2006), 288-301
doi:10.1016/j.nuclphysb.2005.12.007
[arXiv:hep-th/0510171 [hep-th]].

\bibitem{TBA}
A.~B.~Zamolodchikov, Nucl.~Phys. {\bf B342} (1990) 695; 
G.~Arutyunov and S.~Frolov,
JHEP \textbf{05} (2009), 068
[arXiv:0903.0141 [hep-th]];
N.~Gromov, V.~Kazakov and P.~Vieira, Phys.\ Rev.\ Lett.\  {\bf 103} (2009) 131601 [arXiv:0901.3753 [hep-th]];
D.~Bombardelli, D.~Fioravanti and R.~Tateo, J.\ Phys.\ {\bf A42} (2009) 375401 [arXiv:0902.3930 [hep-th]].

\bibitem{Arutyunov:2008zt}
G.~Arutyunov and S.~Frolov,
Nucl. Phys. B \textbf{804} (2008), 90-143
[arXiv:0803.4323 [hep-th]].

\bibitem{glebBound} 
G.~Arutyunov, M.~de Leeuw and A.~Torrielli,
Nucl.\ Phys. {\bf B819} (2009) 319 [arXiv:0902.0183 [hep-th]].

\bibitem{niklasYang}
N.~Beisert, PoS \textbf{SOLVAY} (2006), 002 [arXiv:0704.0400 [nlin.SI]];
M.~de Leeuw,
JHEP \textbf{06} (2008), 085
[arXiv:0804.1047 [hep-th]].

\bibitem{BKV}
B.~Basso, S.~Komatsu and P.~Vieira, arXiv:1505.06745 [hep-th].

\bibitem{cushions}
B.~Eden and A.~Sfondrini,
JHEP {\bf 1710} (2017) 098 [arXiv:1611.05436 [hep-th]].

\bibitem{shotaThiago1}
T.~Fleury and S.~Komatsu,
JHEP {\bf 1701} (2017) 130 [arXiv:1611.05577 [hep-th]].

\bibitem{positivity}
B.~Eden, Y.~Jiang, M.~de Leeuw, T.~Meier, D.~le Plat and A.~Sfondrini,
JHEP \textbf{11} (2018), 097 [arXiv:1806.06051 [hep-th]].

\bibitem{shotaThiago2}
T.~Fleury and S.~Komatsu,
JHEP {\bf 1802} (2018) 177 [arXiv:1711.05327 [hep-th]].

\bibitem{fivePt}
M.~de Leeuw, B.~Eden, D.~l.~Plat and T.~Meier,
arXiv:1907.07014 [hep-th]; 
M.~De Leeuw, B.~Eden, D.~Le Plat, T.~Meier and A.~Sfondrini,
arXiv:1912.12231 [hep-th].

\bibitem{Arutyunov:2009ga}
G.~Arutyunov and S.~Frolov,
J. Phys. A \textbf{42} (2009), 254003
doi:10.1088/1751-8113/42/25/254003
[arXiv:0901.4937 [hep-th]].

\bibitem{Arutyunov:2009ce}
G.~Arutyunov, M.~de Leeuw and A.~Torrielli,
JHEP \textbf{05} (2009), 086

\bibitem{Beisert:2014hya}
N.~Beisert and M.~de Leeuw,
J. Phys. A \textbf{47} (2014), 305201
[arXiv:1401.7691 [math-ph]].

\bibitem{Arutyunov:2009kf}
G.~Arutyunov and S.~Frolov,
J. Phys. A \textbf{42} (2009), 425401
[arXiv:0904.4575 [hep-th]].

\end{thebibliography}
\end{document}